\def\year{2022}\relax
\documentclass[letterpaper]{article} 
\usepackage{aaai22}  
\usepackage{times}  
\usepackage{helvet}  
\usepackage{courier}  
\usepackage[hyphens]{url}  
\usepackage{graphicx} 
\usepackage{natbib}  
\usepackage{caption} 
\DeclareCaptionStyle{ruled}{labelfont=normalfont,labelsep=colon,strut=off} 
\usepackage{enumitem}

\usepackage{algorithm}
\usepackage{algorithmic}

\usepackage{newfloat}
\usepackage{listings}
\floatstyle{ruled}
\newfloat{listing}{tb}{lst}{}
\floatname{listing}{Listing}
\title{AI and Identity}

\author{
    Sri Yash Tadimalla, Mary Lou Maher \textsuperscript{\rm 1}\\
}
\affiliations{
    \textsuperscript{\rm 1}University of North Carolina Charlotte\\

    stadimal@uncc.edu, mmaher9@uncc.edu 
}

\begin{document}

\maketitle

\begin{abstract}

AI-empowered technologies' impact on the world is undeniable, reshaping industries, revolutionizing how humans interact with technology, transforming educational paradigms, and redefining social codes. However, this rapid growth is accompanied by two notable challenges: a lack of diversity within the AI field and a widening AI divide. In this context, This paper examines the intersection of AI and identity as a pathway to understand biases, inequalities, and ethical considerations in AI development and deployment. We present a multifaceted definition of AI identity, which encompasses its creators, applications, and their broader impacts. Understanding AI's identity involves understanding the associations between the individuals involved in AI's development, the technologies produced, and the social, ethical, and psychological implications. After exploring the AI identity ecosystem and its societal dynamics, We propose a framework that highlights the need for diversity in AI across three dimensions: Creators, Creations, and Consequences through the lens of identity. This paper proposes the need for a comprehensive approach to fostering a more inclusive and responsible AI ecosystem through the lens of identity. 

\end{abstract}

\section{Introduction}

With the rise of AI, there's an emergence of complex questions about the very fabric of its existence. Whether it is politicians mulling over policy implications  \cite{coeckelbergh2022political}, civilians navigating the implications of AI in daily life \cite{liu2023citizens}, Educational institution's decisions on AI usage in learning and teaching \cite{mouta2023uncovering}, Tech organizations using AI to innovate and create business applications \cite{bessen2023business} or military organizations strategizing defense \cite{de2017artificial}, each group views AI through its unique lens \cite{calo2017artificial}. The current discourse on Artificial Intelligence (AI) reflects a complex interplay of optimism and concern regarding AI's impact on global equity \cite{crawford2021atlas}. This narrative is increasingly focused on the "AI Divide," a term that encapsulates the challenges AI poses in perpetuating disparities between those who have access and those who don't \cite{carter2020exploring}. 

The global population is currently around 8.1 billion. As of 2023, it's estimated that there were roughly 3.4 billion people employed worldwide, with 2 billion people working within the informal economy \cite{Dyvik_2023}. The ICT sector was projected to employ 55.3 million people full-time by 2020, according to estimates made before the COVID-19 pandemic \cite{Sherif_2023}. Within this sector, a smaller number of workers are actively involved in developing or working with AI tools and systems. Looking forward, it's expected that by 2025, up to 97 million individuals will be working in the AI field. Furthermore, the US AI market is anticipated to reach 299.64 billion dollars by 2026. AI tools and systems are projected to impact nearly 40 percent of jobs globally, with this figure rising to about 60 percent in advanced economies \cite{Howarth_2024}. These changes will lead to a combination of job replacement and augmentation with the influencing factor being access to or lack of access to the internet and digital infrastructure.

This scenario underscores a significant power imbalance, where a relatively small portion of the global population has a profound influence on the lives and livelihoods of the majority as visualized in Figure 1. In this rapidly evolving landscape and widening AI divide, a notable challenge that has emerged is the significant lack of diversity among the creators, researchers, and educators in the AI field. This homogeneity within the AI workforce, if it persists into the predicted 96 million AI workforce, represents more than an issue of fairness or representation; it fundamentally affects the design, implementation, and impact of AI technologies across different populations and societies.  \cite{stathoulopoulos2019gender}.

The concern is that while AI has immense potential for progress in areas like healthcare, agriculture, and education, its development is predominantly driven by the private sector \cite{abboud2020redefining} \cite{chui2023economic}, raising critical questions about its applicability and fairness \cite{dwivedi2021artificial}. Renowned AI safety groups and industry leaders have raised alarms not just about existential risks akin to pandemics and nuclear war but also about the social inequities AI might exacerbate \cite{hinton2023statement} \cite{buolamwini2018gender}. Scholars have highlighted the immediate dangers of AI, especially its impact on marginalized and vulnerable populations \cite{birhane2022forgotten}. This is evidenced by the shortcomings of existing AI models, including facial recognition inaccuracies and the limited effectiveness of Large Language Models (LLMs) in addressing the diverse needs of global populations \cite{coulter2023ai}. Congressional testimony on the need for AI regulation and the collective call by AI researchers for a pause in AI advancements emphasize the urgency of addressing these issues \cite{pause2023experiments}. Moreover, the public's divided perception of AI, akin to debates in cybersecurity \cite{kovavcevic2020factors}, points to a significant gap between the aspirational goals of various AI frameworks and principles, and the reality of their implementation, especially in diverse global contexts\cite{hagerty2019global}.

\begin{figure}
    \centering
    \includegraphics[width= 1\linewidth]{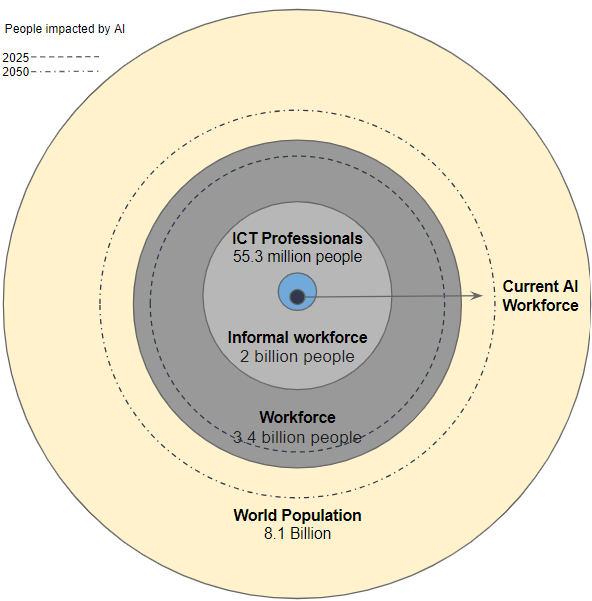}
    \caption{ Representation of global workforce and AI industry impact}
\end{figure}

This evolving narrative and the need to solve the AI diversity gap underscores the need for a nuanced, socio-technical approach to understanding the multifaceted identity of AI \cite{lee2021ai}. Through an exploration of how AI is created, applied, and perceived \cite{kim2023one} across various societal strata, as presented in the AI identity ecosystem (Figure 2), we can answer questions like, Who creates AI? Whose values does it reflect? How is it perceived across different cultures and demographics? And, how do all these factors influence the technology's trajectory?\cite{bingley2023enlarging} It is intricately woven with the myriad societal dynamics that shape identity formation and influence perceptions. The answers to these questions are pivotal as they influence not only the technological trajectories of AI but also its societal implications as presented in the AI identity Framework (Figure 2).  It is critical to underscore that the identity of AI isn't simply a reflection of its perception but a pathway to bridging the AI divide and improving inclusivity efforts.
\begin{figure*}
    \centering
    \includegraphics[width=\textwidth,height=6cm]{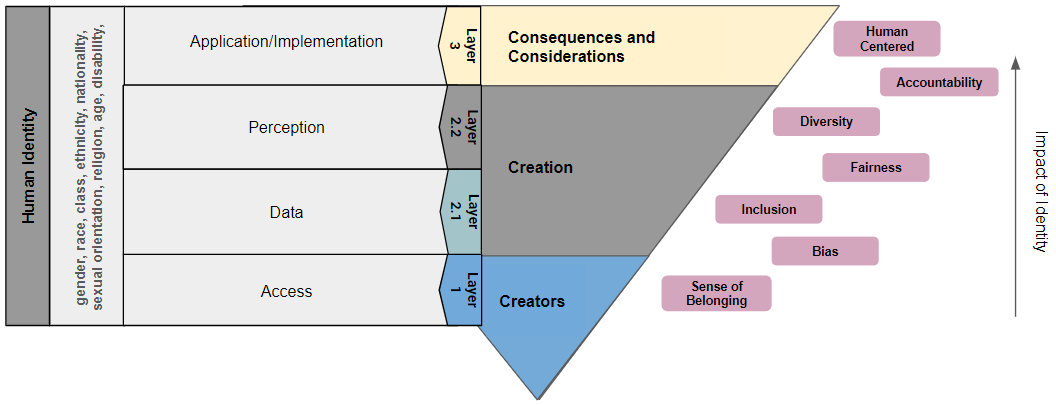}
    \caption{The AI Identity Research Framework}
\end{figure*}

\section{Background}

The process of developing the construct of "AI identity" begins with understanding the drivers and landscape of AI development \cite{christian2020alignment} which can be understood through the AI identity ecosystem (Figure 2). The AI ecosystem is a dynamic and interwoven network comprising a myriad of essential components that collectively drive the evolution and application of artificial intelligence \cite{jacobides2021evolutionary} \cite{yoo2020study}. We will try to explore the various components of the AI ecosystem in this paper. Firstly, the building blocks of the system lay in the technologies and tools, such as machine learning algorithms and frameworks like TensorFlow and PyTorch \cite{paszke2019pytorch}. Complementing these are specialized hardware solutions like GPUs and TPUs, designed to cater to the computational demands of AI \cite{batra2019artificial}. 
Data, the lifeblood of AI, flows in from diverse sources, including IoT devices and online interactions, and is stored and processed using advanced tools \cite{aragon2022human}. 

However, in understanding AI's evolution it is essential to understand the confluence of various human endeavors that drive its development as mentioned above. In this section particularly we focus more on the identity of the creators as it has cascading effect on the other layers. These human contributors do more than just define AI's trajectory; they embed societal norms, values, and biases into the technology, shaping the future of our automated world \cite{cave2020whiteness}. The role of research cannot be overstated, as institutions constantly push the boundaries of what's possible, often dictating the future direction of AI \cite{basole2021visualizing}. Simultaneously, both tech behemoths and startups shape the AI landscape with their innovative products and platforms. Open-source communities bolster this growth, fostering collaboration and ensuring that AI tools remain accessible to all \cite{quan2018understanding}.

As the influence of AI permeates society, regulatory and ethical bodies are stepping in to ensure its deployment is responsible and aligned with societal values \cite{minkkinen2021towards}. End-users and consumers, both individuals and businesses, are the final and important piece of the puzzle, adopting AI solutions and applications and influencing their development through feedback. Integral to all these components and often the ones that shape the direction of development is the financial pillar of investors and funding entities that fuel the growth of AI across sectors \cite{khanna2023progressive}. Lastly, the AI ecosystem is enriched by educational institutions and platforms that impart knowledge, nurturing the current and future generations of AI enthusiasts and professionals through AI literacy \cite{long2020ai} and AI education \cite{schiff2022education}. 

 The identity of AI continues to extend beyond just its creators and the diversity within the field. It encompasses a range of social considerations that shape its existence and impact\cite{cole1991artificial}. Ethical concerns surrounding AI, such as bias in algorithms, privacy issues, and the societal implications of AI, require careful attention and mitigation \cite{amershi2020culture}. Human-AI collaboration is evolving to where AI can augment human capabilities and automate tasks, necessitating the exploration of inclusive ways for humans and AI to collaborate seamlessly \cite{amershi2019guidelines}. The tendency for humans to personify or anthropomorphize AI systems raises psychological and sociological questions about the implications of interacting with AI, especially in education \cite{maher2023exploratory} and work environments \cite{mirbabaie2022rise}. Moreover, public sentiment around AI is often polarized based on users' backgrounds, leading to inaccurate narratives about AI in society\cite{PRAHL2021102077}. These narratives and biases, combined with the digital divide that exists in society and computer science exclude some and inhibit many's access to AI careers, especially for certain groups of underrepresented minorities, and women \cite{shams2023ai}. 

By examining the roles, interactions, and contributions of these diverse stakeholders as presented in the AI Ecosystem (Figure 3) we can gain a comprehensive understanding of how AI systems are created, operate, and evolve. By foregrounding these concepts, we aim to dissect how they influence AI creators and their creations. Through the depiction of layers and various pipelines (connections) across the layers that lead to various consequences or issues in the AI landscape, Figures 1 and 2 together show the myriad ways AI technologies can either reinforce existing societal disparities or potentially pave the way for a more inclusive and equitable future.

By dissecting these relationships, we can better comprehend and address the disproportionate consequences and impacts that AI can have on various segments of society. The construct of human identity is influenced as depicted in the image is present across the layers, present in the identities, data, perceptions, and experiences of end users in the  AI ecosystem. Here Human Identity, as defined within intersectionality \cite{crenshaw2013demarginalizing} and identity theory\cite{jenkins2014social}, encompasses the multifaceted and interconnected aspects of an individual's social categories, including race, gender \cite{scheuerman2020we}, class \cite{inaba2021social}, sexuality \cite{keyes2021truth}, and disability \cite{trewin2019considerations}, nationality and age \cite{pollack2005intelligent} etc, recognizing that these overlapping identities shape unique experiences of privilege and oppression \cite{benjamin2019race}.

\section{AI Identity}

Sociology Literature provides a comprehensive exploration of how individuals' identities are formed \cite{stets2000identity}, negotiated \cite{jenkins2014social}, and transformed \cite{collins2020intersectionality} within the context of social structures and interactions, offering insightful analysis on the complexity of identity as both a personal and social construct \cite{albert1998metadefinition}. This concept also extends to objects, organizations, and technology \cite{wood2004online}, illustrating how identity formation transcends individual experiences to encompass broader societal interactions and constructs \cite{hirsch1992concept}. Often discussions and models involving identity in the AI landscape and ecosystem are heavily centered on the technological and economic aspects \cite{Mitchell_2020}. With efforts focused on defining and explaining what AI is \cite{wang2019defining} and how to understand its usage \cite{devedzic2022identity} \cite{touretzky2023machine}, how AI perceives the human identity \cite{scheuerman2020we} \cite{schlesinger2018let} \cite{tian2017towards}, how humans perceive AI \cite{ragot2020ai} \cite{lima2020collecting} \cite{shinners2022exploring}, how humans interact with the AI  \cite{keyes2021truth} \cite{ashktorab2020human} and how AI influences the human identity in various scenarios \cite{cao2023dark}. This includes discussions on algorithms \cite{noble2018algorithms} \cite{pasquale2015black}, data\cite{aragon2016developing} \cite{aragon2022human}, hardware infrastructure \cite{batra2019artificial}, application areas \cite{huang2021strategic}, market growth, company roles \cite{alahmad2020artificial}, and investment trends \cite{mir2022ai}. While these aspects are undeniably crucial, they often overshadow the deeper, more intricate layers of AI's perception and relationship with the concept of identity.

We define "AI Identity" in two dimensions: internal and external.

\textit{
\textbf{ Internally, AI Identity includes the collective characteristics, values, and ethical considerations embodied in the creation of AI technologies. Externally, AI identity is shaped by individual perception, societal impact, and cultural norms.}
}

These dimensions form a comprehensive view of AI identity, highlighting the interplay between the creation of technology itself and its broader interaction with society. This means understanding the place of AI in society, its development, interactions with individuals \cite{gutoreva2024sharing}, and the nuances of its impact on various facets of human life. The identity of AI is intricately linked to various ethical dilemmas, including responsibility, accountability, fairness, transparency, and trust \cite{benjamin2019race}. These issues are central to the ongoing discussions surrounding the regulation and governance of AI, as well as its cultural and social impacts\cite{arora2023risk}. Furthermore, it is vital to recognize the role of media representations of AI in popular culture, as they significantly shape public attitudes and beliefs about this technology. In this context, the emergence of Human-Centered AI (HCAI) \cite{shneiderman2021human}, with its emphasis on considering human values and agency, represents a pivotal shift in the AI landscape. International organizations like the European Union and research and education institutions are advocating for HCAI, promoting a humanistic and ethical approach that enhances human capabilities while addressing the multifaceted challenges associated with AI identity and its broader societal implications ~\cite{capel2023human}. These dynamic changes in the AI landscape and the important role that education plays in creating opportunities for minorities to participate in AI \cite{su142013572}, led the authors to focus on this topic. In the next three sections of the paper, we look deeper into the layers of identity in AI: the creators, the creations, and the consequences that emerge from the identity ecosystem. Figure 4 summarizes our exploration. By examining the intersection of AI with identity, we aim to shed light on the complex dimensions of AI identity and its implications for individuals and society as a whole.

\begin{figure*}
    \centering
    \includegraphics[width=\textwidth,height= 10cm]{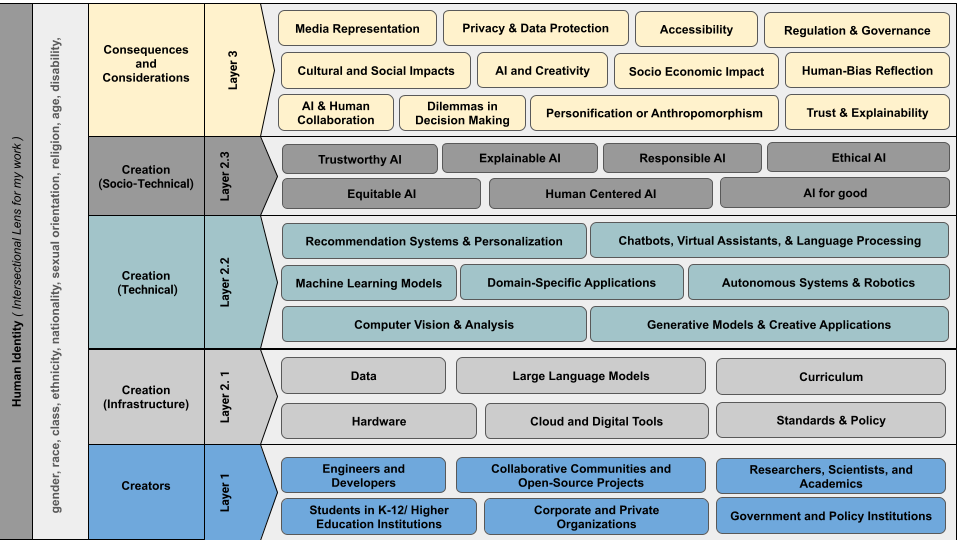}
    \caption{The AI Identity Ecosystem}
\end{figure*}

\section{ The Creators of AI}

The development of AI is influenced by its creators, whose diverse backgrounds and identities are crucial in shaping the technology's direction \cite{schiff2020s}. Despite significant advancements, AI suffers from a notable lack of diversity among its developers, echoing broader inclusivity challenges within the tech industry \cite{inaba2021social}. Diversity is not just about fairness but is essential for creating AI systems that are unbiased, ethical, and beneficial for society \cite{shams2023ai}. Thus, enhancing the diversity of AI creators is imperative to ensure that the technology reflects a broad spectrum of human experiences and values, contributing positively to societal progress\cite{chubb2022expert}. It is important to note that this layer of the ecosystem is closely connected with the internal dimension of AI identity. In the list below, we highlight some of the key groups (individual and organizational) that influence who the creators are, illustrating the nature of AI's development landscape. The goals of AI are often defined by this layer of AI ecosystem and when we look at these key groups, this position paper acknowledges the need to increase diversity across all of the groups across sectors and disciplines.
 
\begin{itemize}[leftmargin=*] 
\item \textbf{Researchers, Scientists, and Academics:} The development of AI is driven by a community of researchers, scientists, and academics from various disciplines \cite{hartmann2020rise}. Experts in computer science, mathematics, biology, linguistics, cognitive science, and engineering collaborate to advance AI technologies. They conduct research, publish academic papers, and contribute to the theoretical foundations of AI\cite{stathoulopoulos2019gender}. Higher education institutions also play a big role in shaping the perception and standards \cite{mouta2023uncovering}that these students carry into their professional careers. With HCI, interaction design, and UX communities contributing to AI work on addressing the gap in discourse about AI's relationship to design practice ~\cite{loi2019co}, higher education institutions and private research institutions are important actors in shaping the diversity in the field of AI \cite{ahmed2020democratization} \cite{whittaker2018ai}. 

\item \textbf{Students and Higher Education Institutions:} Students pursuing higher education, such as undergraduate and graduate programs, play a vital role in the creation and advancement of AI. They learn the fundamental concepts and skills necessary for AI development and contribute fresh perspectives and ideas \cite{touretzky2019envisioning}. Many universities and educational institutions offer specialized AI programs and research opportunities \cite{luckin2016intelligence}, nurturing the next generation of AI creators beginning from K-12 or high school \cite{adejoro2023empower}. It is important to not here the disciplinary differences across AI perception and adoption across various disciplines \cite{zawacki2019systematic}

\item \textbf{Corporate and Private Organizations:} Numerous companies and organizations are actively involved in the development and deployment of AI technologies \cite{jacobides2021evolutionary}. Major tech companies like Google, Microsoft, and IBM invest significant resources into AI research and development \cite{de2021companies}. Startups and research organizations focused on AI innovation also contribute to the creation of novel AI applications and technologies \cite{farber2024analyzing}. Many historically consumer or service-oriented organizations and companies are also building in-house AI solutions across sectors \cite{simon2023ai}.

\item \textbf{Engineers and Developers:} The creation of AI systems requires the expertise of engineers and developers who design and build the underlying software and hardware infrastructure. They develop algorithms, models, and frameworks that power AI applications. These professionals work on coding, testing, and optimizing AI systems to enhance their performance and capabilities \cite{griffin2023ethical}.

\item \textbf{Collaborative Communities and Open-Source Projects:} Collaboration and open-source initiatives play a crucial role in AI creation. Online communities, forums, and platforms facilitate knowledge sharing, collaboration, and the exchange of ideas among AI enthusiasts and professionals \cite{bostrom2018strategic} \cite{hibbard2008open}. Open-source projects like TensorFlow and PyTorch provide accessible frameworks and tools that empower developers to create AI solutions \cite{paszke2019pytorch}. Non-profits and professional societies are also getting involved with building standards and expectations around AI integration from the lenses of ethics and standards. 

\item \textbf{Government and Policy Institutions:} Governments and policy institutions contribute to the creation of AI by establishing frameworks, regulations, and policies that guide its development and deployment \cite{bareis2022talking} \cite{luo2023rise}. The government also invests in the development of AI through its military/defense wing to be up to date with private system developments. Mulitaleral Global organizations in the ICT space are trying to address ethical considerations, and privacy concerns, and ensure responsible and accountable AI practices \cite{vacarelu2022politicians} \cite{floridi2021ethical}. Government-funded National research initiatives and grants also support AI research and innovation \cite{donlon2024national}.

\end{itemize}

\section{The Creation of AI}

 To understand the definition of AI and its application in various scenarios it is important to understand the very building blocks of AI, and how they work and interact with the physical world \cite{raghavan2021societal}. This layer of the ecosystem acts as the mechanism or conduit that translate the goals reflected by the identity of the creators of AI into the consequences and considerations layer of AI Identity ecosystem, and therefore is key to understanding both the external and internal aspects of AI Identity. this should be explained in the beginning of this section and then summarized at the end of this section. In this paper, we will adopt a comprehensive approach to understanding the workings of artificial intelligence (AI) by analyzing it from both technical and socio-technical perspectives. It is simultaneously important to understand the intricate connection between the infrastructure layer and the growth of AI's impact on society through these two lenses.
 
 The technical approach involves understanding \textbf{AI Models and Applications}, where we explore the technical foundations and applications of AI, including chatbots, virtual assistants, language processing, recommendation systems, domain-specific applications, autonomous systems, robotics, machine learning models, computer vision, and generative models. This layer represents the nuts and bolts of AI, where one learns what these technologies are, how they operate, and the specific tasks they are designed to perform. By understanding the mechanics and functionalities of these AI applications, one can appreciate the technical goals and innovations driving AI development. It is important to note here that the development of these goals have historically been  oriented around efficiency and performance not about diversity and inclusion. 
 
 Conversely, the socio-technical approach centers on \textbf{AI Frameworks and Principles}, focusing on the overarching frameworks and principles guiding the ethical, responsible, and equitable development and deployment of AI technologies. This layer encompasses Trustworthy AI, Responsible AI, Ethical AI, AI for Good, Human-Centered AI, and Equitable AI. Here, we assess how these guiding principles influence the design, purpose, and impact of AI technologies on society. This approach allows us to understand the broader implications of AI, including ethical considerations, societal impacts, and the role of AI in promoting social good and addressing global challenges.

This infrastructure layer is not only a precursor to the technical layer's growth but also a fundamental aspect that interweaves with the socio-technical layer, affecting how AI integrates and resonates within human society. Through its capacity to support and scale AI development, It not only provides the technical means for AI's existence but also influences the objectives, ethics, and inclusivity of AI systems by virtue of access. By enabling or limiting the development of certain AI applications, the infrastructure layer shapes the AI identity, reflecting the priorities and values of its creators and users. Furthermore, by offering metrics for capacity building, the infrastructure layer facilitates the assessment of global AI development efforts, guiding policy, investment, and education towards a more equitable and inclusive AI future. 

\subsection{The Infrastructure Layer}
The infrastructure underpinning artificial intelligence (AI) is the bedrock upon which the technical evolution and societal integration of AI is built. This infrastructure encompasses the essential tools, hardware, and frameworks necessary for the development, training, and deployment of AI systems, including data processing capabilities, computational power, and cloud services. These components are indispensable for the processing of vast datasets, supporting complex model training, and ensuring the scalable deployment of AI applications. As such, the infrastructure directly influences the pace and direction of AI innovation. The disparities in AI capabilities across different regions highlight the need for global cooperation and investment in AI infrastructure. Consequently, infrastructure availability becomes a crucial metric for gauging capacity-building efforts, underscoring the importance of democratizing AI benefits to mitigate regional disparities. As AI technologies become increasingly embedded in everyday life, the infrastructure that supports these technologies plays a vital role in determining how they are received, understood, and integrated into various cultural contexts. This interaction between AI and society influences the design, functionality, and adoption of AI systems, necessitating a thoughtful approach to infrastructure development that considers and reflects diverse human values and societal needs.

\begin{itemize}[leftmargin=*]

\item  \textbf{ Large Language Models (LLMs):}
Large Language Models, such as GPT (Generative Pretrained Transformer) and BERT (Bidirectional Encoder Representations from Transformers), represent significant advancements in natural language processing and understanding. These models, which learn from vast amounts of text data, have revolutionized how machines understand and generate human-like text, enabling a wide range of applications from automated writing assistants to advanced chatbots. The development and refinement of LLMs require substantial computational resources and sophisticated algorithms, showcasing the importance of robust infrastructure in AI innovation \cite{devlin2018bert} \cite{brown2020language}.

\item  \textbf{ Data: }
Data serves as the cornerstone for training AI models, providing the raw material from which machines learn and make inferences. The quality, diversity, and volume of data directly impact the performance and bias in AI systems. Effective data management and processing infrastructure are crucial for handling the increasing scale of data, necessitating advanced storage solutions and data processing frameworks \cite{hartmann2020rise}. The ethical considerations in data collection and use are also paramount, emphasizing the need for transparency and fairness in AI \cite{aragon2022human}.

\item  \textbf{ Curriculum:}
The AI curriculum encompasses the educational resources and programs designed to equip individuals with the knowledge and skills required to develop and manage AI technologies. This includes courses on machine learning, ethics in AI, data science, and specialized AI applications. The development of a comprehensive AI curriculum that is accessible and inclusive is essential for fostering a diverse and skilled workforce capable of advancing AI technology while considering its societal impacts \cite{expandcapacity2024} \cite{long2020ai} \cite{schiff2022education} \cite{song2024framework}.

\item  \textbf{ Standards and Policy:}
Standards and policies play a critical role in guiding the ethical development, deployment, and governance of AI technologies. These frameworks help ensure that AI systems are developed responsibly, promoting transparency, accountability, and fairness. International organizations and regulatory bodies are increasingly focusing on developing AI standards and policies that balance innovation with ethical considerations and human rights \cite{jobin2019global} \cite{long2023ai}

\item  \textbf{ Hardware:}
The hardware infrastructure, including GPUs (Graphics Processing Units) and TPUs (Tensor Processing Units), is fundamental for training and running AI models \cite{jouppi2017datacenter}. These specialized processors can handle the parallel processing of large datasets and complex algorithms, making them indispensable for deep learning applications. The development of more efficient and powerful hardware is crucial for advancing AI capabilities and making AI more accessible and sustainable \cite{Merritt_2024b}.

\item  \textbf{ Cloud and Digital Tools:}
Cloud computing and digital tools offer scalable and flexible resources for AI development and deployment. Cloud services provide access to computational power, storage, and AI development tools on-demand, facilitating the rapid prototyping and scaling of AI applications. This infrastructure supports the collaborative development of AI, enabling access to cutting-edge technology without significant upfront investment \cite{armbrust2010view}.

\end{itemize}

In essence, the infrastructure supporting AI is not just a foundation for its technical growth but also a mediator of its interaction with society, influencing both the capabilities of AI technologies and their alignment with human values. 

\subsection{The Technical Layer: AI Models and Applications}
Understanding the foundation models of AI is essential to recognize how data influences AI's identity and its diverse applications, from enhancing daily life to revolutionizing industries. While AI moves beyond the imagery of futuristic robots to practical applications like chatbots, self-driving cars, and medical diagnosis, concerns about inclusivity and a human-centric focus persist. The distinction between Symbolic AI and Connectionist AI, such as deep learning, highlights different approaches to AI development, emphasizing the need for AI to be both transparent and grounded in human values.

As AI's applications extend from machine learning models in image recognition to personalized recommendation systems and beyond, the adaptability and potential of AI are evident. However, aligning AI with societal needs and ethical considerations remains a challenge. The exploration of AI's capabilities across various domains underscores the importance of developing AI technologies that are inclusive, ethical, and beneficial for all, ensuring a positive contribution to society's global challenges. Some of the concepts and applications that one needs to familiarize themselves with to get a comprehensive understanding of foundational AI blocks:

\begin{itemize}[leftmargin=*]

\item  \textbf{Machine Learning Models:} Often categorized into supervised learning, where models learn from labeled data; unsupervised learning, identifying patterns in unlabeled data; and reinforcement learning, where models optimize behavior through rewards from trial and error actions. AI employs machine learning models, including deep neural networks, for tasks like image recognition and natural language processing \cite{goodfellow2016deep}. However, while these models excel at pattern recognition and decision-making based on data, they sometimes lack a human-centered focus, which can result in outputs that might not always align with human values or societal nuances \cite{raghavan2021societal}. 

\item  \textbf{Chatbots, Virtual Assistants, and Language Processing:} AI-powered chatbots, virtual assistants (e.g., Siri, Alexa), and language processing systems leverage natural language understanding and generation to interact with humans, provide information, and perform tasks \cite{khurana2023natural} \cite{kasneci2023chatgpt}.

\item  \textbf{Autonomous Systems and Robotics:} AI enables the development of autonomous systems, including self-driving cars \cite{grigorescu2020survey}, drones, and robots, which can make decisions and perform tasks without direct human intervention \cite{thrun2000probabilistic}.

\item  \textbf{Generative Models and Creative Applications:} AI generative models \cite{devlin2018bert}, like GANs and VAEs, create realistic content such as images, music, and text, blurring the line between human and machine-generated creations \cite{9115874}. AI is used in creative fields\cite{du2020creativity} such as art, music, and literature \cite{SurveyGDM}.

\item  \textbf{Recommendation Systems and Personalization:} AI-powered recommendation systems analyze user behavior and preferences to provide personalized recommendations for various products, services, and content, enhancing user experiences and engagement \cite{LU20121}. Personalized Marketing and Targeted Advertising AI analyzes user data and behavior to deliver personalized marketing strategies, including targeted advertisements and recommendations tailored to individual customers. 

\item  \textbf{Computer Vision and Image Analysis:} AI advancements in computer vision enable machines to understand and interpret visual information, performing tasks such as image recognition, object detection, facial recognition, and autonomous surveillance \citet{iandola2016squeezenet} \cite{voulodimos2018deep}.

\item  \textbf{Domain-Specific Applications:} Medical Diagnosis, Healthcare, and Fraud Detection AI systems aid in medical diagnosis, analyzing patient data, medical images, and clinical records \cite{topol2019deep}. They also contribute to fraud detection and cybersecurity, identifying patterns and anomalies to protect against threats.
 
\end{itemize}

\subsection{The Socio-technical Layer: AI Frameworks and Principles}

The frameworks of Trustworthy AI, Responsible AI, Ethical AI, AI for Good, Human-Centered AI, and Equitable AI originated from the growing awareness of the profound impacts AI technologies have on society, coupled with a recognition of the potential risks and ethical dilemmas they present. As AI began to permeate every aspect of human life, from healthcare and education to governance and privacy, scholars, policymakers, and technologists acknowledged the need for guiding principles to ensure the development and deployment of AI technologies that are beneficial, fair, and aligned with human values \cite{floridi2021ethical}. 

\begin{itemize}[leftmargin=*]
\item  \textbf{Trustworthy AI:} This framework focuses on creating AI systems that are safe, transparent, and reliable, emphasizing user privacy and data security \cite{li2023trustworthy}. It aims to build user trust through transparency, fairness, and accountability \cite{ai2019high}.

\item  \textbf{Responsible AI:} This framework stresses the ethical development and use of AI, ensuring societal benefit and minimal harm. It advocates for ethical standards and the consideration of AI's broader societal impacts, promoting integrity in AI deployment \cite{dignum2019responsible}.

\item  \textbf{Ethical AI:} These AI principles align AI development with ethical principles, including human rights and fairness, to prevent bias and discrimination. It is supported by frameworks like those from the \cite{jobin2019global}, encouraging diverse perspectives in AI development.

\item  \textbf{AI for Good:} This framework applies AI to tackle global challenges \cite{rakova2023algorithms} and achieve the United Nations Sustainable Development Goals \cite{coeckelbergh2021ai}, focusing on healthcare, education, and environmental protection \cite{floridi2021ethical}. It highlights AI's potential for positive societal impact \cite{cowls2019designing}.

\item  \textbf{Human-Centered AI (HCAI):} These AI principles prioritize human needs and values in AI development, integrating psychology, sociology, and ethics to enhance human capabilities without compromising dignity or autonomy. It advocates for participatory design, making AI accessible and meaningful \cite{shneiderman2021human}.

\item \textbf{ Explainable AI:} As AI becomes more complex and powerful, there is a growing need for trust and explainability. Users and stakeholders want to understand how AI systems reach their conclusions or recommendations \cite{arrieta2020explainable}. Research on explainable AI aims to develop methods that can provide meaningful explanations for AI decisions \cite{arrieta2020explainable}.

\item  \textbf{Equitable AI:} These AI principles aim for the fair distribution of AI benefits and the mitigation of harms across society, addressing inequalities in AI's development and deployment. It focuses on diversity, inclusivity, and access, promoting inclusive design practices ~\cite{bennett2020care}.

\end{itemize}

The perception of AI and these frameworks dynamically influence each other\cite{shin2021effects}.
These frameworks play a crucial role in shaping the development of AI by providing ethical, social, and technical guidelines that aim to maximize the benefits of AI while minimizing harm. They serve as a compass for creators, guiding the design of AI systems that respect human rights, promote inclusivity, and ensure accountability. They collectively emphasize ethical standards, societal welfare, and human values, advocating for transparency, fairness, and accountability \cite{floridi2021ethical}. However, their focuses vary slightly; Trustworthy AI prioritizes user trust through safety and reliability, Responsible and Ethical AI emphasizes adherence to ethical standards and societal impacts, AI for Good targets the application of AI for solving global challenges, Human-Centered AI stresses enhancing human capabilities and well-being, and Equitable AI seeks to ensure fairness and prevent exacerbation of inequalities. 

Together, these frameworks form a comprehensive set of principles guiding the ethical, inclusive, and beneficial development of AI technologies, highlighting the multifaceted approach needed to address the complex implications of AI in society. Public perception impacts the emphasis and direction of these frameworks, as societal concerns about privacy, job displacement, and bias in AI systems have led to a greater focus on trustworthiness, responsibility, and ethics in AI development. Conversely, the adoption and promotion of these frameworks can positively influence public perception of AI, building trust and confidence in AI technologies.

\section{AI consequences}

This section explores the second dimension of the AI identity that is external and introduces a wide range of ethical, societal, and cultural considerations. Ethical issues, including algorithmic bias, privacy breaches, and the potential for surveillance, pose significant concerns. The integration of AI into everyday life highlights the potential for synergy between humans and AI, facilitating collaborative problem-solving while also raising questions about anthropomorphism and its implications. The reflection of human biases in AI systems underscores the urgency for fairness and inclusivity in model development. As AI systems become more complex, the demand for transparency and clear explanations of AI decisions grows, alongside the need for stringent regulations and robust data protection measures. AI's foray into creative fields prompts discussions about originality and the nature of human creativity. Moreover, AI's impact extends to reshaping job markets and socioeconomic structures, necessitating strategies to protect workers and leverage AI for socioeconomic improvement. Media representations of AI significantly influence public perception, highlighting the need for a critical analysis of how AI is portrayed and perceived. 

The concept of intersectional identity is pivotal when discussing perception, revealing how AI's interaction with culture, economics, and ethics intersects with various human identities, such as race, gender, socioeconomic status, and disability. Biased AI systems disproportionately affect marginalized communities, amplifying inequalities. The differential impact of AI on job markets, based on intersecting identities, calls for equitable AI deployment and strategies tailored to diverse societal needs. Media representations must also embrace intersectionality, offering nuanced portrayals that challenge stereotypes and shape diverse perceptions of AI technologies. Acknowledging these intersectional dimensions is essential for the development, regulation, and integration of AI technologies in a manner that ensures fairness, inclusivity, and equitable outcomes for everyone.

\begin{itemize}[leftmargin=*]

\item  \textbf{Ethical Considerations:} The development and deployment of AI raise important ethical considerations. Issues like bias in AI algorithms, privacy concerns, job displacement, and the impact of AI on society and human values require careful consideration and mitigation. Beyond its creators, AI's identity is shaped by a plethora of ethical dilemmas that have broader societal implications. The dichotomy of AI applications, as seen in authoritarian regimes using it for surveillance versus democratic societies grappling with its ethical deployment, illustrates the complexity of its societal implications \cite{jobin2019global}.

\item \textbf{Personification or Anthropomorphism:} Humans often tend to anthropomorphize AI systems, attributing human-like qualities, intentions, and emotions to them \cite{cole1991artificial}. This phenomenon raises important questions about the psychological and societal implications of interacting with AI. One of the most vivid examples of humans attributing lifelike qualities to AI systems can be seen in the applications of voice-activated virtual assistants like Siri, Alexa, or Google Assistant \cite{Blut2021}. As benign as this may seem, it brings forth significant questions. How does this humanization of machines affect our social behaviors, especially among the younger generation? Do these interactions blur the lines between genuine human relationships and interactions with coded algorithms? Additionally, as AI systems get more sophisticated in their responses, the risk of over-trusting or becoming emotionally reliant on them grows. This phenomenon necessitates a broader discussion on the implications of AI-human interactions, not just from a technological standpoint, but from psychological and societal perspectives as well.

\item \textbf{AI and Human Collaboration:} The relationship between AI and humans is evolving. AI can augment human capabilities, automate repetitive tasks, and provide valuable insights. Human-AI collaboration is essential for leveraging the strengths of both to tackle complex problems ~\cite{dellermann2019hybrid}but introducing AI may negatively impact employees if not done intentionally \cite{mirbabaie2022rise} \cite{rezwana2023designing}.

\item \textbf{Human-Bias Reflection:} AI systems can inherit human biases present in the data they are trained on, leading to biased outcomes and discriminatory behavior \cite{schellmann2024algorithm}. Recognizing and mitigating these biases is essential to ensure fairness and inclusivity in AI applications \cite{buolamwini2018gender}. We have seen the need to scrutinize and rectify any biases in the data of AI systems become critical especially in areas like criminal justice, to ensure fairness and prevent perpetuation of historic injustices.

\item \textbf{Regulation and Governance:} The rapid advancement of AI has prompted discussions around the need for regulations and governance frameworks \cite{hinton2023statement} \cite{pause2023experiments}. Governments and organizations are exploring ways to ensure the responsible and ethical use of AI, protect privacy, and address potential risks associated with its deployment \cite{madiega2021artificial}.

\item \textbf{Privacy and Data Protection:} AI systems rely on vast amounts of data, raising concerns about privacy and data protection. Experts in computer science and law collaborate to develop frameworks and regulations that safeguard individuals' privacy while allowing for AI innovation \cite{aragon2022human} \cite{arora2023risk}.

\item \textbf{AI and Creativity:} AI is making its mark in creative domains such as art, music, and literature. This blurs the line between human creativity and machine-generated output \cite{karimi2020creative}, raising questions about authorship, originality, and the nature of art. Can an AI, devoid of lived experiences and emotions, genuinely craft a narrative that resonates with human emotions? And if the AI art/creation wins a competition, who would take the accolades - the AI, its developers, or the myriad of authors whose works trained the system? \cite{franceschelli2022copyright}

\item \textbf{Cultural and Social Impacts:} AI technologies are not developed in a vacuum but are shaped by cultural and social contexts. Experts in sociology and anthropology examine how AI systems reflect and reproduce societal norms, values, and power dynamics \cite{karizat2021algorithmic}. 

\item \textbf{Socioeconomic Impact: }The implications of AI-driven automation have the potential to reshape job markets and socioeconomic structures through job displacement and disrupting the human resources landscape \cite{schellmann2024algorithm}.  Thus, exploring strategies to ensure a just transition for workers and leveraging AI for socioeconomic development becomes crucial across various sectors like healthcare, education, and governance \cite{chui2023economic}. Countries that rapidly adapt to AI stand to gain significant socioeconomic advantages, while those lagging behind face an exacerbated digital divide. This widening gap between AI-ready nations and those without adequate access or infrastructure could lead to significant disparities in social and economic growth in the coming decades escpecially for marginalised groups \cite{chetty2014land}.

\item \textbf{Media Representations:} Media plays a crucial role in shaping public perceptions and understandings of AI. Experts critically analyze how AI is portrayed in popular culture, news media, and entertainment, especially the genres of science fiction and speculative fiction, which have a long history of shaping the public's perception of AI. These diverse media representations significantly influence public sentiment and understanding of AI, thus shaping the discourse surrounding its development and deployment in real-world contexts. 

\item \textbf{AI Dilemmas:} Various combinations of Consequences and considerations in AI raise numerous dilemmas, such as determining responsibility and accountability for AI actions, ensuring fairness and transparency in decision-making, and considering the ethical impact of AI on social dynamics, safety, and privacy ~\cite{caldwell2022agile}. Highlighting that we must ensure that while AI is a valuable resource and tool, the final decision must rests with the human expert, capitalizing on the strengths of both entities to address intricate challenges while promoting the principles of HCAI.  

\end{itemize}
 When AI is seen as being developed and deployed in a manner that is aligned with ethical principles and societal well-being, it garners broader acceptance and support ~\cite{lucivero2020data}. This interplay highlights the importance of transparent communication and engagement with the public in the ongoing development and refinement of AI technologies and their guiding frameworks.
 \begin{figure}
    \centering
    \includegraphics[width= 1\linewidth]{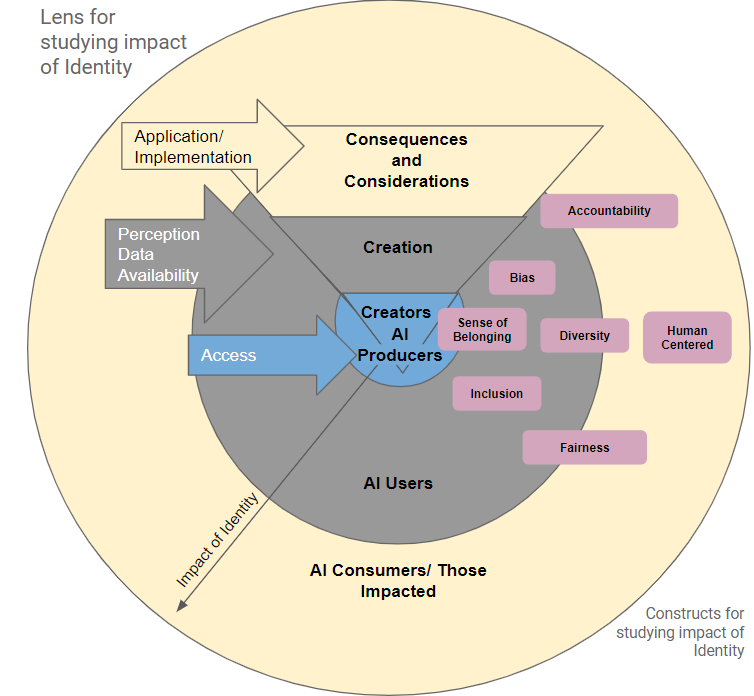}
    \caption{The AI Identity Framework}
\end{figure}

\section { AI Identity Framework}

The perception of AI is not monolithic, It is influenced by numerous factors, including societal structures, personal experiences, media representations, etc. The lens of identity in the field adds a layer of complexity, which leads to discussions about whether AI application design and creation processes are informed by a myriad of backgrounds, experiences, and worldviews. Thus by understanding the context of who creates AI technologies, as well as examining the frameworks and societal considerations that guide AI's development and the way they impact society, we can advocate for diversity and inclusion as essential to an AI Identity that serves all people fairly. The research and conceptual frameworks of AI identity ecosystem capture this complexity by examining the relationship between individual identity factors—such as race, ethnicity, gender, class, sexual orientation, religion, and disability—and their impact in technological contexts, specifically through the lens of creators and consumers roles in the development, access to, data management, perception, and implementation of AI in applications/solutions\cite{kivunja2018distinguishing}. In the technology development landscape, the creator's work is influenced by their experiences, perceptions and identity, which manifests in the data they select and collect, ultimately shaping the technology they create. Thus representation and inclusion in the creation process has far-reaching consequences and considerations, which the framework suggests must be critically assessed through the lens of identity. Key sociological constructs such as diversity, fairness, inclusion, and bias are interwoven with a fundamental sense of belonging, and accountability underscoring the importance of these concepts in evaluating diversity and inclusion work in the field of AI. The upward arrow alongside 'Impact of Identity' suggests that the presence or lack of diversity and inclusion across the mentioned sociological constructs get amplified as we move up through the layers. for example, the impact of bias in the creator's layers \cite{10.1145/3457607} snowballs exponentially into the consequences and considerations layer much akin to the bio accumulation/magnification process in nature \cite{bommasani2022opportunities}. Greater emphasis on these aspects can potentially elevate the role and positive influence of identity in the technological sphere. This sociological framework serves as a guide for a comprehensive analysis of how identity shapes technology and, conversely, how technology can reflect and affect societal values and individual sense of self.

In essence, this position paper underscores the importance of an interdisciplinary, identity-centered approach when educating future designers and developers of AI systems. 
It reinforces the belief that for AI to be truly beneficial for all of society, the designers of tomorrow must be equipped not just with technical expertise, but with a profound understanding of their own identity and understand the ethical, social, and personal ramifications of AI's role in our world. Importantly, this approach combined with Human-Centered principles can also catalyze broadening participation. When AI technologies are built to prioritize inclusivity and fairness, they can naturally spark interest and engagement from a broader cross-section of society. A commitment to center AI creation and development in identity and the human experience can lead to more people, including those from underrepresented backgrounds, feeling motivated to participate in the field. In return, their participation ensures a broader scope of insights, improving AI's responsiveness to societal needs.

\section {Conclusion}
In conclusion, the world of AI is ever-changing, with new creators, creations, and ideas constantly emerging as technology advances. Highlighting the interplay between the technology itself and its broader interaction with society, We define AI Identity in two dimensions to form a comprehensive view of AI identity. AI Identity that includes the collective characteristics, values, and ethical considerations embodied in the creation of AI technologies internally, and AI identity that is shaped by individual perception, societal impact, and cultural norms externally.The discussions within this paper shed light on the significant impact of diversity and inclusion in shaping public perceptions and understanding of AI, demonstrating how these narratives influence the discourse surrounding AI technologies in various societal contexts. Moreover, by proposing the AI identity framework, which captures the impact of various social constructs such as diversity, fairness, inclusion, bias, sense of belonging, and accountability across the creators, creations, and consequences of AI, we advocate for a more inclusive and responsible AI ecosystem. This AI identity ecosystem lens highlights the need for the development of AI technologies that are equitable, accessible, and beneficial for all segments of society.  This paper serves as a call to action, urging the AI community to ground the development of AI in the human experience. An approach that creates technology to addresses the needs of diverse populations, which, in turn, fosters greater inclusivity and engagement in AI development. 


\bibliography{aaaipaper2024}

\bigskip

\end{document}